**Rapid Single-Cell Measurement of Transient Transmembrane Water Flow under Osmotic Gradient**

Hong Jiang[1], Jinnawat Jongkhumkrong,[2] Y. J. Chao[1]; Qian Wang[2*]; Guiren Wang[1,3*]

[1]Department of Mechanical Engineering,

[2]Department of Chemistry and Biochemistry,

[3]Department of Biomedical Engineering,

University of South Carolina, Columbia, SC 29208, USA

* Email: wang263@mailbox.sc.edu (Q. Wang); guirenwang@sc.edu (G. Wang).

While aquaporin (AQP) gating dynamically regulates transmembrane water permeability for cellular homeostasis, its mechanisms remain poorly understood compared to ion channels. A central challenge is the lack of methods to measure water flow through AQPs with the spatiotemporal resolution and sensitivity equivalent to patch-clamp recordings of ion fluxes—a limitation stemming from the electrically silent nature of water transport. We introduce a technique to rapidly detect cytoplasmic flows induced by osmotic-gradient-driven transmembrane water transport in single adherent human cancer cells. This approach enables direct measurement of AQP-mediated water transport and provides a powerful tool to investigate AQP function and regulation and cytoplasmic flow dynamics at the single-cell level.

Transmembrane water transport across cell membranes is essential for maintaining homeostasis and regulating cellular function and volume[1]. Dysfunction of this process contributes to various diseases[2,3] and has been an active research topic for over a century[2,4]. Aquaporins (AQPs), a family of water-selective channels discovered approximately 30 years ago[5,6], are critical for osmotic balance and water homeostasis in organisms ranging from mammals to microbes and plants[7,8]. AQPs' primary function is to facilitate rapid water movement across cytoplasmic membranes in response to osmotic shock[9,10], enabling biological processes, such as cell migration and blebbing[10,11]. AQPs have also been implicated in pathologies including cancer metastasis and Alzheimer's disease[12,13], highlighting their translational potential as targets for diagnostic and therapeutic applications[14,15].

To characterize AQP function and regulation, rapid and direct measurement of transmembrane water flows at single cells with high sensitivity is essential. However, unlike ion channels—where ion flux can be quantified as an electric current via patch clamp technique— AQPs are electrically silent and water flow lacks an electrical signal[14,16,17]. As a result, no method currently matches the sensitivity of electrophysiology for measuring water transport under osmotic gradients applied near the cell surface. Instead, cell volume change rate is typically measured in response to osmotic shock to estimate transmembrane water permeability ($P_m$), a key metric for evaluating AQP function [7,18]. Various methods have been developed to determine $P_m$ by tracking cell volume changes or solute concentration shifts caused by the volume change during osmotic challenge [7,19,20]. While these techniques have advanced our understanding of transmembrane water transport for characterizing AQPs' function, they have low temporal resolution, sensitivity and are limited by experimental artifacts and intrinsic challenges in measuring $P_m$ accurately [9,20,21].

Furthermore, in physiological contexts, cell volume does not always change significantly under an osmotic gradient. For example, during water reabsorption across tight urinary epithelia—driven by a transepithelial osmotic gradient under antidiuretic hormone control—the cell volume in the collecting tubules of the rabbit kidney cortex increases by only ~2.3% during transcellular water flow[22]. Similarly, endothelial cells in microvessels experience a nearly steady osmotic gradient and flux from the lumen to the interstitium, as described by Starling's law[23,24]. In such cases, transcellular convection occurs without substantial changes in cell volume. Accurately measuring volume changes and plasma membrane surface area in these adhered cells is challenging and detecting transmembrane water transport signals requires new methodologies.

In the absence of hydrostatic pressure difference between the extracellular medium and cytoplasm, traditional methods estimate the osmotic water permeability $P_m$ by measuring change rates in cell volume. The rate of volume change is given by

$$\frac{dV_c}{dt} = -P_m A_c V_w (c_{im} - c_{ic}) = -P_m A_c V_w \Delta c \tag{1}$$

where $V_c$ is the cell volume, t time, $V_w$ the partial molar volume of water, $A_c$ the membrane cross-sectional area across which water flows into or out of the cell, $c_{im}$ and $c_{ic}$ the osmolality of impermeant solutes on the medium and cytoplasm sides, $\Delta c = c_{im} - c_{ic}$ the osmolality difference at the membrane interface[25-27].

For our approach, Eq. (1) can be rearranged to express the cytoplasmic bulk flow velocity as:

$$U_c = \frac{dV_c}{A_c dt} = -P_m V_w \Delta c \tag{2}$$

where $U_c$ is the bulk flow velocity in cytoplasm (or its deviation from the initial condition $U_c = 0$) induced by transmembrane water flow driven by an osmotic gradient $\Delta c$ cross $A_c$. According to Eq. (2), if $U_c$ can be measured instantaneously for a known $\Delta c$, then, there is no longer a need to determine $A_c$ or wait for a measurable cell volume change to estimate $P_m$. Notably, both $P_m$ and $U_c$ share the same unit, suggesting that, in addition to $P_m$, $U_c$ could serve as a novel biophysical marker for characterizing AQPs' function in response to rapid osmotic perturbations $\Delta c \neq 0$, independent of volume change. In general, measuring $U_c$ as an instantaneous quantity is significantly faster than tracking the cumulative change in $V_c$.

Cytoplasmic flows in large cells have been measured using nanoparticles ($\geq$ 23 nm diameters) as tracers, whose volume is approximately $\sim 10^5$ times larger than that of water molecules[28,29]. However, the 20-40 nm pores of the filamentous meshwork, combined with cytoplasmic organelles and crowders, limit particle mobility and compromise flow velocity measurements when using tracer particles[30,31]. By contrast, small molecule fluorescent dyes enable precise detection of extremely low flow velocities (~1 µm/s) induced by osmotic shock. These velocities are predicted by Eq. (3) for a large membrane permeability coefficient $P_m$ = 0.1 mm/s [19] under a 300 mOsm gradient ($\Delta c$) generated by DI water as the hypotonic solution[32]. Here, we demonstrate that a flow-induced fluorescence increase velocimeter (FIFIV), which employs small molecular dye rather than nanoparticles, can rapidly capture transient cytoplasmic flow signals ($U_c$) triggered by a localized osmotic shock $\Delta c \neq 0$ in single cells. FIFIV is based on a laser-induced fluorescence photobleaching anemometer (LIFPA), which offers high spatial and temporal resolution[33-35].

Fig. 1a illustrates the generation of local cytoplasmic flow $U_c$ under an imposed osmotic shock $\Delta c \neq 0$. The cell is initially equilibrated in an isotonic medium (e.g., 1× PBS buffer). A localized pulse injection of a hypotonic solution ($-\Delta c > 0$; typically, a DI water droplet) near the left membrane establishes an instantaneous asymmetric osmotic gradient across the cell, while the right side remains exposed to near-isotonic conditions. This drives transient left-to-right transmembrane water flow, subsequently inducing a cytoplasmic flow. If FIFIV can rapidly detect the cytoplasmic velocity change $U_c$, then, according to Eq. (2), $P_m$ can be determined in real time without requiring measurements of cell volume and surface area—parameters that are difficult to obtain rapidly for adherent cells with irregular geometry.

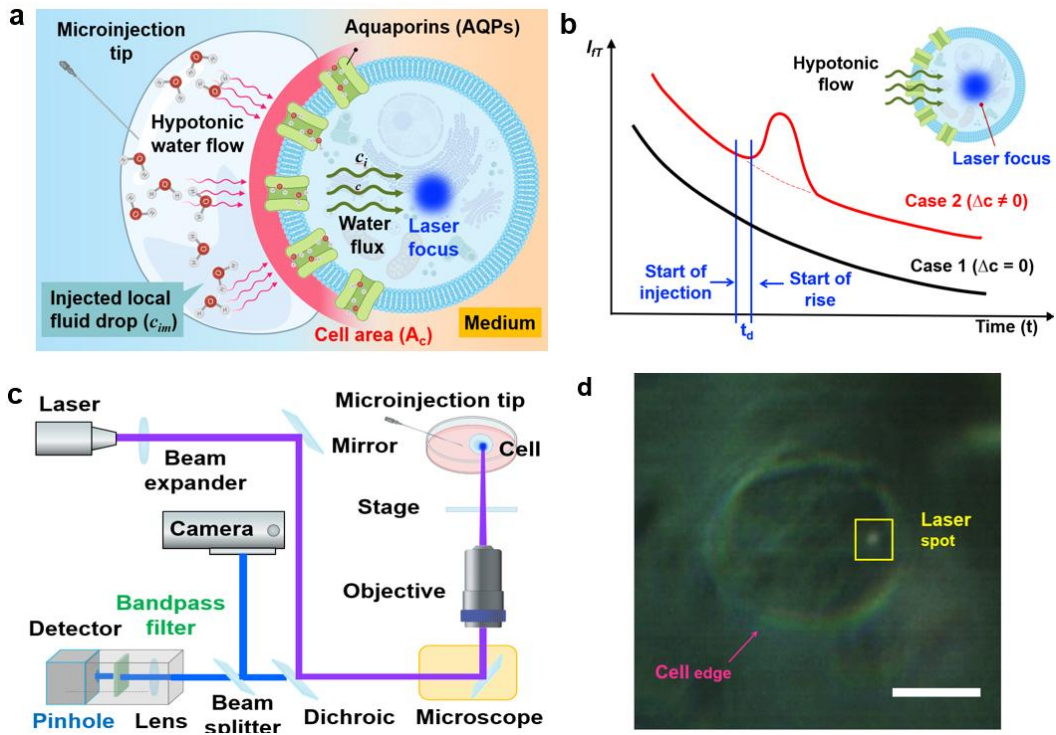

Fig. 1 Principle and setup of the FIFIV system. a. Schematic of transmembrane water flow induced by $\Delta c \neq 0$, generated by injection of a DI water droplet on the left side of a cell, while the right side remains exposed to the original isotonic medium. b. Principle of FIFIV: Case 1 shows the baseline condition $\Delta c = 0$; Case 2 illustrates the presence of $\Delta c \neq 0$. c. Experimental setup for measuring cytoplasmic velocity $U_c$. d. Representative image showing the laser focal point aligned within the cytoplasm of a single cell. The scale bar: 10 μm.

The principle of using FIFIV to measure $U_c$ is illustrated in Fig. 1b through two representative cases. In Case 1 ($\Delta c = 0$), when a laser beam is focused within the quiescent cytoplasm of an adhered cell at rest, the fluorescence signal $I_{fT}$ decays exponentially due to photobleaching (black curve). In contrast, Case 2 depicts the response to an osmotic shock: as shown in Fig. 1a, a localized injection of a drop of hypotonic solution on the left side of the cell establishes an osmotic shock ($\Delta c \neq 0$), causing a transmembrane water flow from left to right. Owing to the incompressibility of cytoplasm, this results in immediate cytoplasmic flow at the focal detection point. Unbleached dye between the focal region and the left membrane is advected into the focal volume, replacing bleached dye and producing an immediate rise in $I_{fT}$. This response follows the LIFPA principle, where the fluorescence signal increases with flow velocity $U_c$ according to: $I_{fT} = I_0 e^{-d_f/(U_c \tau)}$. Here, $I_0$ is the initial $I_{fT}$, $d_f$ the focal spot diameter, and $\tau$ the half decay constant[33]. This assumes constant dye concentration outside the focal volume over short timescales. Once osmotic equilibrium is reached, the flow ceases, resulting in a peak of $I_{fT}$, followed by another decay as shown in case 2 (red curve) in Fig. 1b. Here, there is a time delay $t_d$ between the moment of osmotic shock injection and the onset of $I_{fT}$ increase as confirmed in Fig. 2a.

A schematic of the FIFIV setup is shown in Fig. 1c, integrated with a Nikon C2 confocal microscope. A 405 nm Cube diode laser (Coherent) was used for fluorescence excitation. The laser beam was expanded with a beam expander to 5 mm in diameter before entering a 60X oil-immersion objective with numerical aperture (NA) of 1.4, achieving near-diffraction-limited spatial resolution of ~ 200 nm laterally and 500-600 nm axially. The laser power at the microscope entrance was modulated to remain below 10 μW at 100 kHz and a pulse width of 190 ns. A PInano XYZ P-545.3C7 Piezo stage (Physik Instrument) with 1 nm resolution was mounted on the microscope to precisely manipulate and align the laser focal point and cell position. For the present experiments, the focal point was inside the cell and positioned 16 µm away from the left cell membrane as shown in Fig. 1d. Fluorescently labeled cells cultured and adhered to the bottom of an imaging dish, were mounted on the stage holder. A collecting lens and optical bandpass filter were placed in front of a multimode optical fiber (core diameter 10 µm) acting as a pinhole to enhance resolution, which was connected to a single photon detection module id100-MMF50 (Becker & Hickl Inc.).

Breast cancer cell line MDA-MB-231, which predominantly expresses AQP1, AQP3, and AQP5[10] was used for the FIFIV measurement of cytoplasmic flows to detect the transmembrane water flows driven by $\Delta c \neq 0$. The cells were cultured in an imaging dish (Celltreat) for over 24 hours prior to testing to ensure firm adhesion to the substrate. The cytoplasm of the live cells was labeled with the small fluorescent dye Calcein AM 450 (Invitrogen), and the dish containing the labeled adhered cells was filled with 1X PBS buffer for experiments of the osmotic shock injection. The final dye concentration in the working solution was 1 µM.

An Eppendorf microinjection system was installed on the microscope for injecting DI water to generate hypotonic $\Delta c \neq 0$, that drives transmembrane water flows across the cell membrane. A CellTram Oil microinjector was used to manually inject DI water drops. The injector was connected to a quartz capillary filled with DI water, with inner and outer diameters of 20 µm and 90 µm, respectively. The injection tip was controlled by a module PiezoXpert and positioned approximately 30 µm from the left side of the cell membrane and 20 µm above the bottom of the dish as illustrated in Fig. 1a. A camera was used to monitor the alignment of the cell, laser focal point and injection tip.

The capability of FIFIV to measure cytoplasmic flows induced by transmembrane flows following a pulse injection of $\Delta c \neq 0$ is demonstrated in Fig. 2, where the laser was activated at approximately 20 s. To confirm that the observed signal arises exclusively from $\Delta c \neq 0$, two control experiments were performed. In the first control, no water was injected at all, resulting in an expected exponential decay of $I_{fT}$ (green curve). In the second, 100 nL of 1X PBS buffer was injected, creating an isotonic solution with $\Delta c = 0$, which also produced a simple exponential decay (blue curve, Fig. 2a) as anticipated. In both control cases, no transmembrane water flow or cell volume change was observed during the measurement. In contrast, when 100 nL DI water was injected, the medium on the left side of the cell became hypotonic, generating a transmembrane flow. Within about 1 second, the exponential decay was interrupted, and the signal began to increase to a local peak after about 6 seconds, followed by a faster decay and then a quasi-exponential decay again, as the transmembrane flow ceased once osmotic equilibrium was reached through diffusion of the droplet with the buffer (red curve, Fig. 2a). The area under the red pulse peak correlates with the total transmembrane water flux induced by $\Delta c \neq 0$ at the single cell level. Thus, the local increase in $I_{fT}$ within the pulse is directly attributable to the imposed osmotic perturbation.

Although fluorescence quenching is often applied in cell shrinking assays[36], the signal increase observed in Fig. 2 cannot be attributed to the decrease in dye concentration resulting from the increased cell volume. This conclusion is supported by our observation that $I_{fT}$ in cytoplasmic images increases with dye concentration in the range of 0.2 - 2 µM following incubation, and the working dye concentration here was approximately 1 µM. Moreover, cell volume changes were nearly negligible during the first 2 s, while $I_{fT}$ had already begun to rise. The slower cell volume increase occurs because, unlike suspension cells where the entire membrane experiences $\Delta c \neq 0$ uniformly, adhered cells respond to localized hypotonic stimulation (DI water droplet on one side), resulting in gradual swelling. The vertical line in Fig. 2a marks the start of the injection. The corresponding cell volume increase 3 min after injection is shown in Fig. 2b and 2c, where the area analysis indicates a volume increase of approximately 6.4%, assuming cell height remained unchanged.

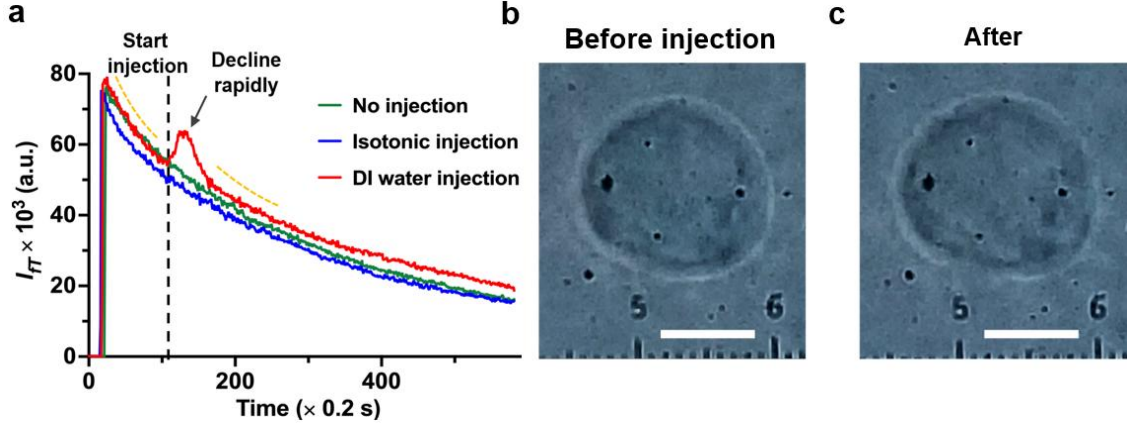

Fig. 2 $I_{fT}$ time series and cell images. (a) $I_{fT}$ response under three conditions: no injection (green); injection of an isotonic solution ($\Delta c = 0$, blue); and injection of a hypotonic solution ($-\Delta c > 0$, red) near the cell surface. In the hypotonic case ( red curve), the exponential decay of $I_{fT}$ is immediately disrupted after the injection, producing a pulse signal corresponding to intracellular flow induced by transmembrane water flow across the cell membrane. (b) Image of the cell before hypotonic injection. (c) Image of the same cell 3 min after injection. Cell volume was increased by approximately 6.4% following the injection, confirming the transmembrane water flow. Scale bar: 10 μm.

Fig. 2a could also be used to estimate $U_c$ and $P_m$. After reaching its maximum, $I_{fT}$ is expected to gradually decrease. However, when water flows into the cell and reaches the focal point, the local dye concentration decreases, causing $I_{fT}$ to decline more rapidly than it would if the dye concentration remained constant. The flow velocity $U_c$ can be approximated as the ratio of the distance between the focal point and the membrane on the injection side to the time interval between the start of the injection and the moment when $I_{fT}$ starts to sharply decrease after the peak, i.e. the time taken for water to travel from the membrane surface to the focal point. For the red curve in Fig 2a, $U_c$ is estimated to be approximately 1.95 μm/s. Using Eq. (2) and assuming $\Delta c$ = 300 mOsm[32], the corresponding $P_m$ is estimated to be ~ 0.0361 cm/s, which is about 60% higher than the reported value of 0.0225 cm/s for the same cell line[37].

Because FIFIV is highly sensitive to flow dynamics, it can also detect the flow changes induced by a rapid injection of $\Delta c \neq 0$ even when cell volume changes are not measurable, as shown in Fig. 3. Fourteen seconds after the laser beam was switched on and focused inside the cell, 160 nL of DI water was injected as shown by the vertical line in Fig. 3a. An immediate rise in $I_{fT}$ was observed within around 1 s, indicating the onset of the transmembrane and cytoplasmic flows generated by the imposed $\Delta c \neq 0$. A comparison of Fig. 3d (before injection) and Fig. 3e (3 min after the first injection) shows an about 12% increase in cell volume. Two minutes later, another 160 nL of DI water was injected, resulting in another $I_{fT}$ increase, as shown in Fig. 3b. However, by 3 min post-injection, the cell volume in Fig. 3f had increased by only 2% compared to that in Fig. 3e. A third injection of 160 nL of DI water conducted 2 min later, still produced a clear $I_{fT}$ rise as shown in Fig. 3c, despite the already elevated cell volume compared to Fig. 3d. Notably, almost no additional volume change was detected in Fig. 3g relative to Fig. 3f, suggesting that the cell volume had reached a temporary quasi-steady state. The signal increase in Fig. 3c may result from a flow behavior analogous to the transcellular flow in endothelial cells in capillaries, where water enters the cell from one side and exits from the other, or from a volume regulation mechanism. The reduced signal amplitudes in Fig. 3b and 3c compared to that in Fig. 3a are attributed to photobleaching and dilution of the fluorescent dye in the cytoplasm due to volume increase. Fig. 3 shows FIFIV could potentially measure transcellular flows as well.

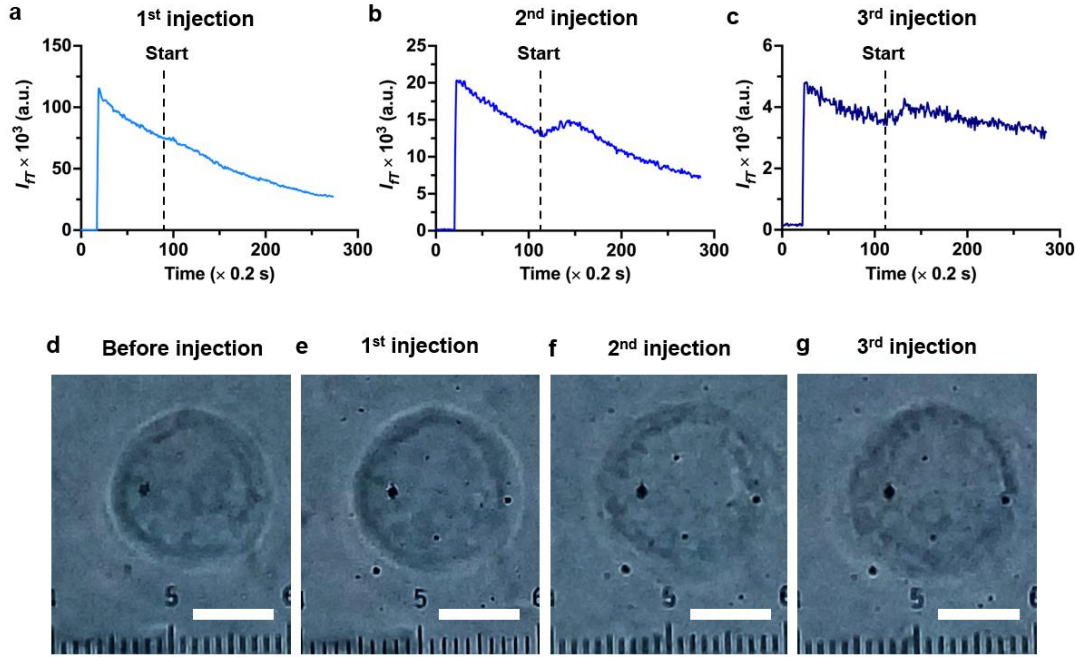

Fig. 3 $I_{fT}$ time series and images of a single cell subjected to multiple DI water injections. a-c: $I_{fT}$ responses to the first, second, and third injections of DI water, respectively, showing pulse signals induced by Δc ≠ 0. d: Image of the cell before the first injection. e-g: Images of the same cell 3 min after the first, second, and third injections, respectively. While clear pulse signals were observed for each injection, the cell volume exhibited almost no additional change after the third injection. This suggests the involvement of a volume-regulation mechanism that limits cell swell further and may allow water to flow out of the cell from the opposite side. Scale bar: 10 μm.

In summary, while transmembrane water flow driven by osmotic gradients lacks electrical signals, the resultant cytoplasmic flow generates a measurable fluorescence signal. Leveraging fluorescence spectroscopy, FIFIV detects instantaneous changes in transmembrane water transport and the induced cytoplasmic flows with high temporal resolution, eliminating the need for slow volume change measurements. This represents a potential paradigm shift for investigating AQP gating mechanisms. FIFIV could characterize AQPs' function and regulation at a single cell level by comparing cells with and without siRNA knockdown of AQPs or by applying AQP modulators under osmotic shock—analogous to how patch-clamp techniques are used to study ion flux through ion channels under a membrane voltage. FIFIV requires no genetic modification and may be directly applied to human epithelial cells in adhered conditions, offering physiological relevance and potential compatibility with high-throughput drug screening and in vitro studies. In addition, FIFIV has the potential to quantify transcellular water transport across microvascular endothelia.

Acknowledgments—We are grateful to Alan Verkman for valuable comments.